\documentclass[12pt]{article}

\usepackage{listings}

\usepackage{epsfig}
\usepackage{epsf}
\usepackage{amsmath}
\usepackage{amsfonts}
\usepackage{amsthm}
\usepackage{amssymb}
\usepackage{enumerate}

\usepackage{pstricks}
\usepackage{lastpage}

\def\C{\mathbb{C}}
\def\R{\mathbb{R}}
\def\i{\bf i}

\usepackage{xspace}
\usepackage{tikz} 
\usepackage{calc} 
\usepackage{ifthen} 

\usetikzlibrary{decorations.pathmorphing, decorations.pathreplacing, patterns, shapes.geometric} 

\tikzset{
    slope/.code={\edef\slope{#1}},
    slope/.default=0.5,
    slope
}
\makeatletter
\pgfdeclarepatternformonly[\tikz@pattern@color,\slope]{slant lines}
{\pgfpoint{-.1mm/\slope}{-.1mm}}
{\pgfpoint{1.1mm/\slope}{1.1mm}}
{\pgfpoint{1mm/\slope}{1mm}}
{
    \pgfsetlinewidth{0.4pt}
    \pgfpathmoveto{\pgfpoint{-.1mm/\slope}{-.1mm}}
    \pgfpathlineto{\pgfpoint{1.1mm/\slope}{1.1mm}}
    \pgfsetstrokecolor{\tikz@pattern@color}
    \pgfusepath{stroke}
}
\makeatother

\newcommand{\myref}[3]{%
  \draw (#1, #2) node {#3};
  \draw[line width = 0.07cm] (#1-0.5, #2-0.6) -- (#1-0.3, #2-0.3) -- (#1+0.3, #2-0.3);
}

\newcommand{\figA}[1]{{\bf A.{#1}}\xspace}
\newcommand{\figS}[1]{{\bf S.{#1}}\xspace}
\newcommand{\figM}[1]{{\bf M.{#1}}\xspace}

 \def\pacase {north east lines}
 
 \def\papcb  {grid}

 \def\paplug {dots}
 
 \def\pamag  {crosshatch dots}

\def\myskip{\smallskip\noindent}

\usepackage{fancyhdr} 
\def\myskip{\smallskip\smallskip\noindent}

\begin{document}

\baselineskip=18pt

\begin{center}
{\large\bf

THE ELEGANT NMR SPECTROMETER

}

\smallskip

Elena Ibragimova, Ilgis Ibragimov

\smallskip

{\it

Elegant Mathematics LLC, 82834 WY USA \&

Elegant Mathematics Ltd, 66564 Germany

\smallskip

e-mail: ii@elegant-mathematics.com

}

\end{center}

{\bf\large

\noindent
INTRODUCTION

}

\myskip
This paper pertains to the field of nuclear magnetic resonance (NMR) spectroscopic techniques and their applications in real-time chemical analysis.

\myskip
One of the biggest disadvantages of low-field NMR spectrometers is the high fluctuation of their magnetic fields. If the magnets are small (about the size of a
portable device), the intensity of the external magnetic field and its direction may be adversely affected. Even turning a 1.5~T NMR spectrometer to an angle
about 6$^{\rm o}$ degrees perpendicular to the earth's magnetic force lines will ruin the measurements, and the device will have to be recalibrated. Even a slight
movement of the table on which a currently available spectrometer is placed may disturb the spectra significantly \cite{NMR-Review}. Another related difficulty
is that currently available spectrometers usually require high temperature stability (of the order of 0.01$^{\rm o}$C), which is incompatible with chemical
production equipment and in-situ measurements in chemical reactions \cite{LFNMR1, LFNMR2}. 

\myskip
There is well-known and widely used approach that improve the sensitivity of NMR measurements, i.e., multi-nuclear and multi-dimensional
spectra acquisition \cite{Nature2006, JACS2008}.

\myskip
The acquisition of multi-nuclear spectra usually requires one receiver coil for each type of nuclei \cite{NMR-Review}, and this fact also makes
currently available NMR spectrometers incompatible with smaller, portable devices. 

\myskip
Hence, prospective inventors of a portable NMR spectrometer for industrial environments must overcome the following problems: 
\begin{itemize}
\item construct a signal acquisition scheme that is stable despite the fluctuations of the permanent magnetic field;
\item construct a signal acquisition scheme that is stable despite the fluctuations of the signal generator or that can work without a signal generator; and
\item use NMR to detect all (or most) visible, non-zero-spin isotopes that exist in the investigated area. 
\end{itemize}

\newpage

{\bf\large

\noindent
ELEGANT NMR

}

\myskip
{\bf E}nhanced multi-nuc{\bf LE}ar {\bf G}eneration, {\bf A}cquisition, and {\bf N}umerical {\bf T}reatment of {\bf N}u\-clear {\bf M}agnetic {\bf R}esonance
spectra ({\bf ELEGANT NMR}) is a processing method constructed according to the following scheme. 

\myskip
Consider FIG. 1: One or several wide-band coils and/or optical detectors \figS2 receive very weak signals that are usually amplified by one or more sequential amplifiers
\figS3. The signals are abbreviated as $f_l(t)$, $l=1,\dots,L$, where $L$ is the total count of input signals and $t$ is the time domain variable of the measurements.
Based on {\it a priori~} information about the magnets and non-zero-spin isotopes in use, one or several frequency generators \figS5 and their signals, delayed on $1/4$ period,
are used. Signals from said frequency generators are abbreviated as $v_n(t), ~~ n=1,\dots,N$, so that $v_n(t)$ is a complex function whose real part refers to the signal
and whose imaginary part refers to the delayed signal from the same generator. 

\myskip
All $f_l(t), l=1,\dots,L$ signals are forwarded pairwise with $v_n=1,\dots,N$ to the mixer block \figS6, so that each pair is comprised of one $f$ and one $s$ signal.
The resulting signals from each mixer are forwarded over a low-pass filter \figS1 and abbreviated as $u_{ln}(t)$, where $l$ is the index of the input NMR coil and $n$ is
the index of the frequency generator. 

\myskip
This method is nowadays well-known and used in many NMR devices; however, the following key differences to prior-art methods are suggested: 
\begin{itemize}
\item all generators $v_n(t), n=1,\dots,N$ are fully correlated to each other, i.e. at any time the frequencies of all generated signals have fixed ratios to one another; 
\item at least two repetitions of data acquisition should be performed; 
\item all NMR input coils perform measurements of the same substance/mixture; 
\item at least one non-zero-spin isotope with {\it a priori~} known spectra and concentration should be either: 
\begin{itemize}
\item situated in the measured substance, or 
\item incorporated as the reference unit inside one or several input coils, or 
\item incorporated in the walls of the measuring NMR camera.
\end{itemize}
\end{itemize}
\myskip
Consider that the $j$-th index in $u_{lnj}(t), j=1,\dots,J$ refers to the number of experiments that are collected in different time-slots as is demonstrated in FIG. 2,
taking into the account that $u_{lnj}(t) \in \C$ is a complex function by construction. 

\myskip
Consider that $q_{ln}(t)$ is the {\it a priori~} known spectrum for a particular input coil and a particular non-zero-spin isotope. In the case that no reference isotope
is present in one or several input coils, the corresponding function $q_{ln}(t)$ is equal to zero. 

\myskip
Since all $v_n(t)$ are fully correlated to each other, magnetic field fluctuations and resonator fluctuations add to each other; this may be further represented as
the fluctuation function $1+\sigma_j(t)$, where $|\sigma_j(t)|<<1$. By construction, $\sigma_j(t)$ differs along the $j$-th index, since fluctuations of both the magnetic
field and the resonator are random. 

\myskip
Hence, the reconstruction of the pure spectrum $p_n(t)$ of the measured mixture is performed by the minimization: 
$$
\min_{p_n(t), \epsilon_j(t)} \sum_{n=1}^N \sum_{l=1}^L \sum_{j=1}^J ||u_{lnj}(t) - (p_n(t) + q_{ln}(t)) \epsilon_j(t)||_2^2, \eqno (1)
$$
where $\displaystyle \epsilon_j(t) = e^{\i W_n \sigma_j(t)}$. The corresponding algorithm that performs this reconstruction is described in Appendix of this paper.

\myskip
There are situations when some entries of $u_{lnj}(t)$ may be missing, for example due to optimization of component count. In these cases, the minimization problem may be
solved with a sparse multi-dimensional approximation according to the algorithms on \cite{IIV2002}.

\myskip
Every signal in $f$, $v$, and $u$ in the described method may be analog or digital. At any point between the blocks \figS1-\figS3, \figS5-\figS9, one or several analog
to digital converters (ADCs) and/or one or several digital to analog converters (DACs) can be incorporated to convert between signal types. Any of the blocks \figS5-\figS8,
can be implemented through analog and/or digital means. In each particular case, the use of digital, analog, or a mix of digital and analog signals is dependent upon component
counts, costs, accuracy, average signal frequency, and many other factors. 

\myskip
$|u_{ln}(t)|$ on the output of \figS1 at FIG. 1 refers to $|p_{ln}(t)|$ and is weakly dependent on fluctuations in the permanent magnetic field and oscillator,
is generated with several microsecond delays after the initial signal appeared, so all real-time methods that require only $|p_{ln}(t)|$ may be used.

\myskip
Compact and portable in-situ NMR spectrometers (FIG. 3) which can be dipped in the liquid to be measured, and are easily maintained, with affordable coil constructions (FIG. 5),
together with an apparatus to recover depleted magnets (FIG. 4) are presented, that provide a new real-time processing method for NMR spectrum acquisition, that remains stable
despite magnetic field fluctuations.

{\bf\large

\noindent
FIGURES

}

\myskip
\underline{\bf FIG. 1}:
A processing method to convert wide-band NMR signals.

\myskip
\underline{\bf FIG. 2}:
A processing method to convert a set of continuously measured experiments delivering $u_{lnj}(t)$ from one set of NMR receivers by collecting several measurements
and solving a minimization (1) using computational unit \figS9.

\myskip
\figS1 Low-pass filter block that is used in parallel with all passed signals.

\myskip
\figS2 Receiver coil.

\myskip
\figS3 One or more sequentially-connected amplifiers.

\myskip
\figS4 A marker --- a substance/mixture containing at least one non-zero-spin isotope with {\it a priori~} known spectra and concentration --- which is either: 
\vspace*{-2mm}
\begin{itemize}
\item situated in the measured substance, or 
\item incorporated as the reference unit inside coils \figS2, or 
\item incorporated in the walls of the measuring NMR camera.
\end{itemize}
\noindent
\figS5 One or several frequency generators and their signals, delayed on $1/4$ period. Each frequency generator has fixed ratio ($a_n/b_n$) to the main frequency generator.

\myskip
\figS6 A set of mixer pairs with each mixer pair receiving a pair of signals

\newpage

\noindent
$(f_l(t), {\rm Re}(v_n(t)))$ or $(f_l(t), {\rm Im}(v_n(t)))$ and delivering their products.

\myskip
\figS7 A processing block that incorporates the method described at FIG. 3.

\myskip
\figS8 A block that continuously supplies pipeline data $u_{lnj}(t)$, $l=1,\dots,L$, $n=1,\dots, N$ from \figS7 into local storage and delivers it to processing
block \figS9.

\myskip
\figS9 A processing block that solves minimization problem (1).


\noindent
\underline{\bf FIG. 3}:
ELEGANT NMR spectrometer embodiment for in-situ measurements, where FIG. 3A refers to the complete assembly, FIG. 3B refers to the component containing the electronics,
FIG. 3C refers to the sensor block for performing measurements in a liquid flow, FIG. 3D refers to the sensor block when dipped in liquid to be measured, and FIG. 3E
demonstrates how the dipped sensor may be constructed to be suitable for standard ground glass joints.

\myskip
\figA1 Data and power supply connector.

\myskip
\figA2 Thermostat connectors: cooling/heating water/liquid/gas is dispersed over these connectors to control the temperature of the electronics inside the device.

\myskip
\figA3 PCB assembly with transmitter and receiver electronics, in case if thermostat connection is used, PCBs should be coated with appropriate materials
to prevent PCB's damage by water/liquid/gas thermostatting.

\myskip
\figA4 Area inside the ELEGANT NMR spectrometer with constant temperature controlled by water/liquid/gas thermostatting.

\myskip
\figA5 Main case of the device for the electronics \figA3.

\myskip
\figA6 Plugs that connect the coils \figA{12} to the electronics \figA3.

\myskip
\figA7 Screw threads to screw the block with magnets \figA{10} to the block with electronics \figA5.

\myskip
\figA8 Gasket for hermetic connection.

\myskip
\figA9 Permanent magnet(s).

\myskip
\figA{10} The case for the block with magnets.

\myskip
\figA{11} Wires to connect plug(s) \figA6 with coil(s) \figA{12}.

\myskip
\figA{12} Receiver and transmitter coils assembly, one embodiment of coils assembly described in FIG. 5. 

\myskip
\figA{13} Flow connectors. One can connect tubes to perform measurements in flow. In the case of the solid-state and NMR tube detector embodiment being used, 
said connectors should allow an external tube to be placed inside the measurement area and associated coils.

\myskip
\figA{14} Ground glass joint.

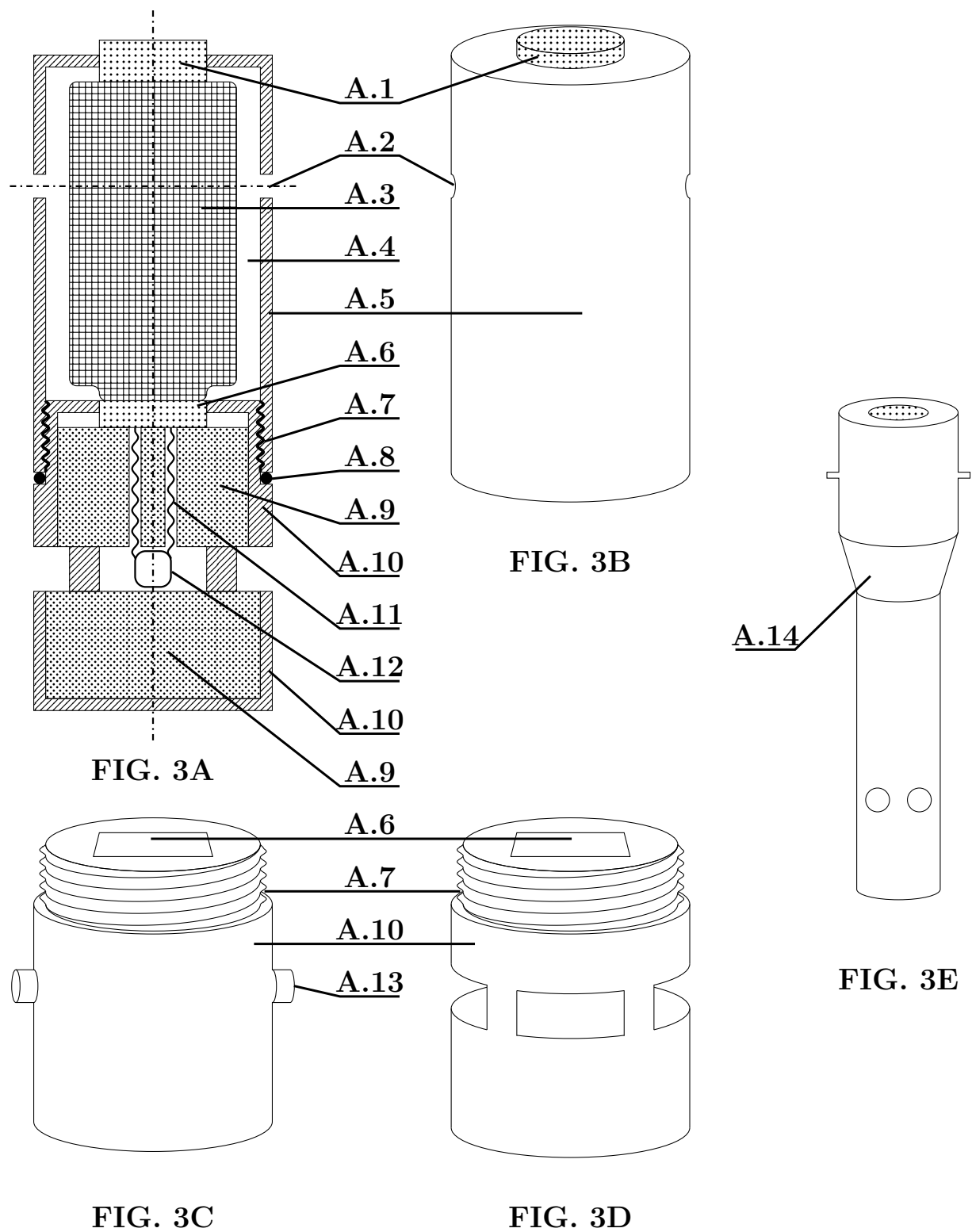
\renewcommand{\figA}[1]{{\bf\large A.{#1}}\xspace}
\renewcommand{\figS}[1]{{\bf\large S.{#1}}\xspace}
\renewcommand{\figM}[1]{{\bf\large M.{#1}}\xspace}

\newpage

\newcommand{\refpointZ}[0]{%
  \draw[line width = 0.04cm] (-2, -2) -- (2, 2);
  \draw[line width = 0.04cm] (-2, 2) -- (2, -2);
}

\def\refpointscale{1.0}

\newcommand{\refpointR}[5]{%
\pgfmathsetmacro\refpointax{(#2-\locoffx)*1.27*\refpointscale/36.}
\pgfmathsetmacro\refpointay{(#3-\locoffy)*1.27*\refpointscale/36.}
\pgfmathsetmacro\refpointbx{(#4-\locoffx)*1.27*\refpointscale/36.}
\pgfmathsetmacro\refpointby{(#5-\locoffy)*1.27*\refpointscale/36.}
\pgfmathsetmacro\refpointsh{(0.5 * \refpointscale}
\pgfmathsetmacro\refpointdx{(0.25 * \refpointscale}
  \draw[line width = 0.04cm] (\refpointax, \refpointay) -- (\refpointbx-\refpointsh, \refpointby) -- (\refpointbx+\refpointsh, \refpointby);
  \draw (\refpointbx, \refpointby+\refpointdx) node {#1};
}

\newcommand{\refpointL}[5]{%
\pgfmathsetmacro\refpointax{(#2-\locoffx)*1.27*\refpointscale/36.}
\pgfmathsetmacro\refpointay{(#3-\locoffy)*1.27*\refpointscale/36.}
\pgfmathsetmacro\refpointbx{(#4-\locoffx)*1.27*\refpointscale/36.}
\pgfmathsetmacro\refpointby{(#5-\locoffy)*1.27*\refpointscale/36.}
\pgfmathsetmacro\refpointsh{(0.5 * \refpointscale}
\pgfmathsetmacro\refpointdx{(0.25 * \refpointscale}
  \draw[line width = 0.04cm] (\refpointax, \refpointay) -- (\refpointbx+\refpointsh, \refpointby) -- (\refpointbx-\refpointsh, \refpointby);
  \draw (\refpointbx, \refpointby+\refpointdx) node {#1};
}

\newcommand{\refpointLR}[7]{%
\pgfmathsetmacro\refpointax{(#2-\locoffx)*1.27*\refpointscale/36.}
\pgfmathsetmacro\refpointay{(#3-\locoffy)*1.27*\refpointscale/36.}
\pgfmathsetmacro\refpointbx{(#4-\locoffx)*1.27*\refpointscale/36.}
\pgfmathsetmacro\refpointby{(#5-\locoffy)*1.27*\refpointscale/36.}
\pgfmathsetmacro\refpointcx{(#6-\locoffx)*1.27*\refpointscale/36.}
\pgfmathsetmacro\refpointcy{(#7-\locoffy)*1.27*\refpointscale/36.}
\pgfmathsetmacro\refpointsh{(0.5 * \refpointscale}
\pgfmathsetmacro\refpointdx{(0.25 * \refpointscale}
  \draw[line width = 0.04cm] (\refpointax, \refpointay) -- (\refpointbx-\refpointsh, \refpointby) -- (\refpointbx+\refpointsh, \refpointby) -- (\refpointcx, \refpointcy);
  \draw (\refpointbx, \refpointby+\refpointdx) node {#1};
}

\vspace*{-15mm}
\noindent
\begin{tikzpicture}

  \def\qq{0.9}

  \draw[line width = 0.04cm, rounded corners=0.2cm] (-5.0*\qq, 4.0*\qq) rectangle (5.0*\qq, 10.0*\qq);
  \myref{5.0*\qq+0.4}{10.0*\qq+0.6}{\figS5}

  \draw[line width = 0.04cm] (0.0*\qq,  9.2*\qq) circle (0.6cm);
  \draw[line width = 0.04cm] (0.0*\qq,  9.2*\qq) arc (0:180:0.25cm);
  \draw[line width = 0.04cm] (0.0*\qq,  9.2*\qq) arc (180:360:0.25cm);

  \draw[line width = 0.04cm] (-4.0*\qq,  4.8*\qq) circle (0.6cm);
  \draw[line width = 0.04cm] (-4.0*\qq,  4.8*\qq) arc (0:180:0.25cm);
  \draw[line width = 0.04cm] (-4.0*\qq,  4.8*\qq) arc (180:360:0.25cm);

  \draw[line width = 0.04cm] ( 0.0*\qq,  4.8*\qq) node {\Large $\dots$};

  \draw[line width = 0.04cm] ( 4.0*\qq,  4.8*\qq) circle (0.6cm);
  \draw[line width = 0.04cm] ( 4.0*\qq,  4.8*\qq) arc (0:180:0.25cm);
  \draw[line width = 0.04cm] ( 4.0*\qq,  4.8*\qq) arc (180:360:0.25cm);

  \draw[line width = 0.04cm] (-4.0*\qq,  6.0*\qq) node {\Large $a_1/b_1$};
  \draw[line width = 0.04cm] ( 0.0*\qq,  6.0*\qq) node {\Large $\dots$};
  \draw[line width = 0.04cm] ( 4.0*\qq,  6.0*\qq) node {\Large $a_N/b_N$};

  \draw[->, line width = 0.04cm] (-0.66*\qq,  9.2*\qq) arc (90:180:3cm and 2.5cm);
  \draw[->, line width = 0.04cm] ( 0.66*\qq,  9.2*\qq) arc (90:0:3cm and 2.5cm);
  \draw[->, line width = 0.04cm] ( 0.00*\qq,  8.54*\qq) -- (0.0*\qq, 6.4*\qq);

  \draw[line width = 0.04cm] (-2.0*\qq,  7.6*\qq) node {\Large $\dots$};
  \draw[line width = 0.04cm] ( 2.0*\qq,  7.6*\qq) node {\Large $\dots$};

  \draw[line width = 0.15cm] (-9.0*\qq, -2.4*\qq) -- (-9.0*\qq, -1.4*\qq);
  \draw[line width = 0.07cm] (-9.0*\qq, -2.4*\qq) -- (-8.6*\qq, -2.8*\qq) -- (-7.8*\qq, -2.8*\qq);
  \draw (-8.2*\qq, -2.4*\qq) node {\figS4};

  \draw[line width = 0.04cm, decorate,decoration={coil,amplitude=.1cm, segment length=0.217cm}] (-9.0*\qq,  0.8*\qq) -- (-9.0*\qq,  2.0*\qq);
  \draw[line width = 0.04cm, decorate,decoration={coil,amplitude=.1cm, segment length=0.217cm}] (-9.0*\qq, -2.0*\qq) -- (-9.0*\qq, -0.8*\qq);
  \myref{-9.0*\qq+0.4}{2.0*\qq+0.6}{\figS2}
  \myref{-9.0*\qq+0.4}{-0.8*\qq+0.6}{\figS2}

  \draw[->, line width = 0.04cm] (-9.0*\qq,  2.0*\qq) -- (-7.4*\qq,  2.0*\qq);
  \draw[->, line width = 0.04cm] (-9.0*\qq,  0.8*\qq) -- (-7.4*\qq,  0.8*\qq);

  \draw[->, line width = 0.04cm] (-9.0*\qq, -2.0*\qq) -- (-7.4*\qq, -2.0*\qq);
  \draw[->, line width = 0.04cm] (-9.0*\qq, -0.8*\qq) -- (-7.4*\qq, -0.8*\qq);

  \draw (-5.7*\qq, 1.4*\qq) node {\large AMPLIFIER};
  \draw[line width = 0.04cm, rounded corners=0.2cm] (-7.4*\qq, 0.4*\qq) rectangle (-4.0*\qq, 2.4*\qq);
  \myref{-4.0*\qq+0.4}{2.4*\qq+0.6}{\figS3}

  \draw (-5.7*\qq, -1.4*\qq) node {\large AMPLIFIER};
  \draw[line width = 0.04cm, rounded corners=0.2cm] (-7.4*\qq, -0.4*\qq) rectangle (-4.0*\qq, -2.4*\qq);
%
%
  \draw[line width = 0.04cm, rounded corners=0.2cm] (-2*\qq, -2*\qq) rectangle (2*\qq, 2*\qq);

  \myref{2*\qq+0.4}{2*\qq+0.6}{\figS6}

\foreach \mmcx in {-\qq, \qq}
{ \foreach \mmcy in {-\qq, \qq}
  { \draw[->, line width = 0.04cm] (\mmcx-0.6, \mmcy) -- (\mmcx-0.20, \mmcy);
    \draw[->, line width = 0.04cm] (\mmcx, \mmcy+0.6) -- (\mmcx, \mmcy+0.20);
    \draw (\mmcx, \mmcy) node {\large $\otimes$};
  }
}

  \draw (-\qq, 0.2*\qq) node {$\vdots$};
  \draw (-0.2*\qq, \qq) node {$\dots$};
  \draw (-0.2*\qq, -\qq) node {$\dots$};

  \draw[->, line width = 0.04cm] (-4.0*\qq, 1.4*\qq) -- (-2*\qq, 1.4*\qq);
  \draw[->, line width = 0.04cm] (-4.0*\qq, -1.4*\qq) -- (-2*\qq, -1.4*\qq);
  \draw (-3.0*\qq,  0.8*\qq) node {\LARGE $f_1(t)$};
  \draw (-3.0*\qq,  0.1*\qq) node {\LARGE $\vdots$};
  \draw (-3.0*\qq, -0.8*\qq) node {\LARGE $f_L(t)$};

  \draw[->, line width = 0.04cm] (-0.6*\qq,  4.0*\qq) -- (-0.6*\qq,  2*\qq);
  \draw[->, line width = 0.04cm] ( 0.6*\qq,  4.0*\qq) -- ( 0.6*\qq,  2*\qq);
  \draw (-1.5*\qq,  3.4*\qq) node {\LARGE $v_1(t)$};
  \draw ( 0.1*\qq,  3.4*\qq) node {\LARGE $\dots$};
  \draw (+1.5*\qq,  3.4*\qq) node {\LARGE $v_N(t)$};

  \draw[->, line width = 0.04cm] (2*\qq,  1.5*\qq) -- (3*\qq,  1.5*\qq);
  \draw[->, line width = 0.04cm] (2*\qq, -1.5*\qq) -- (3*\qq, -1.5*\qq);

  \draw[line width = 0.04cm, rounded corners=0.2cm] (3*\qq, -2.0*\qq) rectangle (5.2*\qq, 2.0*\qq);
  \draw (4.1*\qq, 0) node {\large $\begin{tabular}{c} LOW \\ PASS \\ FILTER \end{tabular}$};
 
  \myref{5.2*\qq+0.4}{2.0*\qq+0.6}{\figS1}

  \draw[->, line width = 0.04cm] (5.2*\qq,  1.5*\qq) -- (8*\qq,  1.5*\qq);
  \draw[->, line width = 0.04cm] (5.2*\qq, -1.5*\qq) -- (8*\qq, -1.5*\qq);
  \draw (6.6*\qq, 1.9*\qq) node {\LARGE $u_{11}(t)$};
  \draw (6.6*\qq, 0.0*\qq) node {\LARGE $\vdots$};
  \draw (6.6*\qq, -1.9*\qq) node {\LARGE $u_{LN}(t)$};

  \draw (0.0*\qq, -3.0*\qq) node {\underline{\bf\large FIG. 1.}};
\end{tikzpicture}


 \newcommand{\drawblock}[2]{%
  \def\yy{#1}
  \draw[line width = 0.04cm, rounded corners=0.2cm] (0.5*\qq, \yy+1.1*\qq) rectangle (5*\qq, \yy-1.1*\qq);
  \draw (2.75*\qq, \yy) node {\large $\begin{tabular}{c} ELECTRICAL \\ CIRCUITS OF \\ FIG. 1 \end{tabular}$};
  \draw[->, line width = 0.04cm] (5*\qq, \yy+0.2*\qq) -- (6.0*\qq, \yy+0.2*\qq);
  \draw[->, line width = 0.04cm] (5*\qq, \yy+0.0*\qq) -- (6.0*\qq, \yy+0.0*\qq);
  \draw[->, line width = 0.04cm] (5*\qq, \yy-0.2*\qq) -- (6.0*\qq, \yy-0.2*\qq);
  \draw (9.5*\qq, \yy) node {\LARGE #2};
  \myref{5*\qq+0.4}{\yy+1.1*\qq+0.6}{\figS7}
 }

\noindent
\begin{tikzpicture}
 \def\qq{0.9}

 \drawblock{5.4*\qq}{                  $u_{111}(t),\dots,u_{LN1}(t)$}
 \draw (9.5*\qq, 3.8*\qq) node {\LARGE $u_{112}(t),\dots,u_{LN2}(t)$};
 \draw (9.5*\qq, 0.0*\qq) node {\LARGE $u_{11J}(t),\dots,u_{LNJ}(t)$};

 \myref{13*\qq+0.4}{6.4*\qq+0.6}{\figS8}

 \draw[->, line width = 0.04cm] (13.0*\qq, 5.4*\qq) arc (90:-90:0.35 and 0.65);
 \draw[->, line width = 0.04cm] (13.0*\qq, 3.8*\qq) arc (90:-90:0.35 and 0.65);
 \draw[->, line width = 0.04cm] (13.0*\qq, 1.5*\qq) arc (90:-90:0.35 and 0.65);

 \draw[->, line width = 0.04cm] (13.7*\qq, 5.4*\qq) -- (13.7*\qq, 0.0*\qq);
 \draw (15.0*\qq, 3.0*\qq) node [rotate=270] {\large $\begin{tabular}{c} SEVERAL \\ REPEATED \\ MEASUREMENTS \end{tabular}$};

 \draw ( 7.5*\qq, 1.9*\qq) node {$\vdots$};
 \draw (11.0*\qq, 1.9*\qq) node {$\vdots$};

 \draw[->, line width = 0.04cm] ( 8.5*\qq, -1.0*\qq) -- ( 8.5*\qq, -2.5*\qq);
 \draw[->, line width = 0.04cm] ( 9.5*\qq, -1.0*\qq) -- ( 9.5*\qq, -2.5*\qq);
 \draw[->, line width = 0.04cm] (10.5*\qq, -1.0*\qq) -- (10.5*\qq, -2.5*\qq);

  \draw[line width = 0.04cm, rounded corners=0.2cm] (7.0*\qq, -2.5*\qq) rectangle (12.0*\qq, -5*\qq);
  \draw (9.5*\qq, -3.75*\qq) node {\large $\begin{tabular}{c} FPGA/CPLD/ \\ ASIC/DSP/ \\ CPU/MCU/GPU \end{tabular}$};

  \myref{12.0*\qq+0.4}{-2.5*\qq+0.6}{\figS9}

  \draw[->, line width = 0.04cm] (7.0*\qq, -3.25*\qq) -- (5*\qq, -3.25*\qq);
  \draw[->, line width = 0.04cm] (7.0*\qq, -3.75*\qq) -- (5*\qq, -3.75*\qq);
  \draw[->, line width = 0.04cm] (7.0*\qq, -4.25*\qq) -- (5*\qq, -4.25*\qq);

  \draw (2.5*\qq, -3.75*\qq) node {\LARGE $p_1(t), \dots, p_n(t)$};

  \draw (14.0*\qq, -4.7*\qq) node {\underline{\bf\large FIG. 2.}};
\end{tikzpicture}
\newpage

\noindent
\begin{tikzpicture}
       \def\sh{7.0cm}   
       \def\sshh{262}    
       \def\upa{0.5cm}  
       \def\sss{0.5cm}  
       \def\ang{45.5}   
       \def\angg{26.8}  

       \def\xe{-0.2cm}  
       \def\xf{-0.3cm}  
       \def\xd{-0.4cm}  
       \def\xc{-0.9cm}  
       \def\xg{-1.4cm}  
       \def\xb{-1.8cm}  
       \def\xa{-2.0cm}  

       \def\yn{11.25cm} 
       \def\yc{11.0cm}  
       \def\yb{10.8cm}  
       \def\ym{10.55cm} 
       \def\ya{9.0cm}   
       \def\yo{8.8cm}   
       \def\yd{8.6cm}   
       \def\yl{5.45cm}  
       \def\ye{5.2cm}   
       \def\yf{5.0cm}   
       \def\yk{4.75cm}  
       \def\ygg{4.0cm}  %
       \def\yg{2.75cm}  
       \def\yh{2.0cm}   
       \def\yi{0.2cm}   
       \def\yj{0.0cm}   

\draw[pattern=\pacase] (\xa, \ya) -- (\xa, \yc) -- (\xc, \yc) -- (\xc, \yb) -- (\xb, \yb) -- (\xb, \ya) -- cycle;
\draw[pattern=\pacase] (-\xa, \ya) -- (-\xa, \yc) -- (-\xc, \yc) -- (-\xc, \yb) -- (-\xb, \yb) -- (-\xb, \ya) -- cycle;
\draw[pattern=\pacase] (\xa, \yj) -- (\xa, \yh) -- (\xb, \yh) -- (\xb, \yi) -- (-\xb, \yi) -- (-\xb, \yh) -- (-\xa, \yh) -- (-\xa, \yj) -- cycle;

\draw[pattern=\pacase] (\xa, \ygg) -- (\xa, \yd) -- (\xb, \yd) -- (\xb, \ye) -- (\xc, \ye) -- (\xc, \yf) -- (\xb+0.2cm, \yf) -- (\xb+0.2cm, \yg) -- (\xa, \yg) --
      (\xa, \ygg-0.2cm) -- (\xb, \ygg-0.2cm) -- (\xb, \ygg) -- cycle;

\draw[pattern=\pacase] (-\xa, \ygg) -- (-\xa, \yd) -- (-\xb, \yd) -- (-\xb, \ye) -- (-\xc, \ye) -- (-\xc, \yf) -- (-\xb-0.2cm, \yf) -- (-\xb-0.2cm, \yg) -- (-\xa, \yg) --
      (-\xa, \ygg-0.2cm) -- (-\xb, \ygg-0.2cm) -- (-\xb, \ygg) -- cycle;

\draw [line width = 0.05cm, decorate,decoration={snake, amplitude=0.05cm, segment length=0.2cm }] (\xb, \ye) -- (\xb, \ygg);
\draw [line width = 0.05cm, decorate,decoration={snake, amplitude=0.05cm, segment length=0.2cm }] (-\xb, \ye) -- (-\xb, \ygg);

\draw [line width = 0.05cm, fill=black] (\xb-0.1cm, \ygg-0.1cm) circle (0.075cm);
\draw [line width = 0.05cm, fill=black] (-\xb+0.1cm, \ygg-0.1cm) circle (0.075cm);

\draw[pattern=\pamag] (\xb, \yh) rectangle (-\xb, \yi);
\draw[pattern=\pamag] (\xe, \yg) rectangle (-\xe, \yk);
\draw[pattern=\pamag] (\xb+0.2cm, \yg) rectangle (\xd, \yk);
\draw[pattern=\pamag] (-\xb-0.2cm, \yg) rectangle (-\xd, \yk);
\draw[pattern=\paplug] (\xc, \ye) rectangle (-\xc, \yk);
\draw[pattern=\paplug] (\xc, \yn) rectangle (-\xc, \ym);
\draw[line width = 0.03cm, rounded corners=5pt] (\xf, \yg-0.075cm) rectangle (-\xf, \yh+0.075cm);
\draw[line width = 0.03cm, decorate,decoration={snake, amplitude=0.05cm, segment length=0.35cm}] ( \xf, \yg-0.2cm) -- ( \xf, \yk);
\draw[line width = 0.03cm, decorate,decoration={snake, amplitude=0.05cm, segment length=0.35cm}] (-\xf, \yg-0.2cm) -- (-\xf, \yk);
\draw[pattern=\papcb, rounded corners=0.12cm] (\xg, \ym) -- (\xg, \yl) -- (\xc, \yl) -- (\xc, \ye) -- (-\xc, \ye) -- (-\xc, \yl) -- (-\xg, \yl) -- (-\xg, \ym) -- cycle;
\draw[pattern=\pacase] (\xg, \yg) rectangle (\xc, \yh);
\draw[pattern=\pacase] (-\xg, \yg) rectangle (-\xc, \yh);
\draw[line width = 0.03cm, dashdotted] (0, \yn+0.5cm) -- (0, \yj-0.5cm);
\draw[line width = 0.03cm, dashdotted] (\xa-0.4cm, \yo) -- (-\xa+0.4cm, \yo);

\draw (0, -1.0cm) node {\bf\large FIG. 3A};

\draw [pattern=\paplug] (\sh+\xc, \yn) -- (\sh+\xc, \yc) arc (180:360:0.9 and 0.225) -- (\sh-\xc, \yn) arc (360:180:0.9 and 0.225) -- cycle;
\draw [pattern=\paplug] (\sh, \yn) ellipse (0.9 and 0.225);
\draw (\sh, \yc) ellipse (2.0 and 0.5);
\draw (\sh+\xa, \yc) -- (\sh+\xa, \ya) arc (90:-90:0.07 and 0.2) -- (\sh+\xa, \ygg) arc (180:360:2.0 and 0.5) -- (\sh-\xa, \yd) arc (270:90:0.07 and 0.2) -- (\sh-\xa, \yc);

\draw (\sh, \ygg-1.5cm) node {\bf\large FIG. 3B};


\def\yy{-7.0cm}

\draw (\sh+\xa, \yy+\yj) arc (180:360:2.0 and 0.5);
\draw (\sh-\xa, \yy+\yh) arc (  0:   -\ang:2.0 and 0.5);
\draw (\sh+\xa, \yy+\yh) arc (180:180+\ang:2.0 and 0.5);
\draw (\sh-\xa, \yy+\yh) arc (  0:    \ang:2.0 and 0.5);
\draw (\sh+\xa, \yy+\yh) arc (180:180-\ang:2.0 and 0.5);
\draw (\sh-\xa, \yy+\yg) arc (  0:   -\ang:2.0 and 0.5);
\draw (\sh+\xa, \yy+\yg) arc (180:180+\ang:2.0 and 0.5);
\draw (\sh+\xa, \yy+\yj) -- (\sh+\xa, \yy+\yh);
\draw (\sh-\xa, \yy+\yj) -- (\sh-\xa, \yy+\yh);

\draw (\sh+\xg, \yy+\yg-0.35cm) -- (\sh+\xg, \yy+\yh-0.36cm);
\draw (\sh-\xg, \yy+\yg-0.35cm) -- (\sh-\xg, \yy+\yh-0.36cm);
\draw (\sh+\xc, \yy+\yg-0.45cm) -- (\sh+\xc, \yy+\yh-0.46cm);
\draw (\sh-\xc, \yy+\yg-0.45cm) -- (\sh-\xc, \yy+\yh-0.46cm);
\draw (\sh+\xc, \yy+\yh-0.45cm) arc (270-\angg: 270+\angg:2.0 and 0.5);
\draw (\sh+\xc, \yy+\yg-0.45cm) arc (270-\angg: 270+\angg:2.0 and 0.5);

\draw (\sh+\xa, \yy+\yg) -- (\sh+\xa, \yy+\yg+1.0cm);
\draw (\sh-\xa, \yy+\yg) -- (\sh-\xa, \yy+\yg+1.0cm);

\draw (\sh, \yy+\yg+1.0cm) ellipse (2.0 and 0.5);

\draw [fill=white, black!0] (\sh+\xa+0.2cm, \yy+\yg+2.0cm) arc (180:360:1.8 and 0.45) -- (\sh-\xa-0.2cm, \yy+\yg+1.0cm) arc (360:180:1.8 and 0.45) -- cycle;
\draw (\sh+\xa+0.2cm, \yy+\yg+2.0cm) arc (180:360:1.8 and 0.45);
\draw (\sh-\xa-0.2cm, \yy+\yg+1.0cm) arc (360:180:1.8 and 0.45);
\draw (\sh, \yy+\yg+2.0cm) ellipse (1.8 and 0.45);
\draw (\sh-1.0cm, \yy+\yg+1.8cm) -- (\sh-0.9cm, \yy+\yg+2.2cm) -- (\sh+0.9cm, \yy+\yg+2.2cm) -- (\sh+1.0cm, \yy+\yg+1.8cm) -- cycle;

\draw (\sh-\xa-0.20cm, \yy+\yg+1.0cm) -- (\sh-\xa-0.15cm, \yy+\yg+1.04cm);
\draw (\sh-\xa-0.20cm, \yy+\yg+2.0cm) -- (\sh-\xa-0.15cm, \yy+\yg+1.95cm);
\draw [decorate,decoration={snake, amplitude=0.05cm, segment length=0.2cm }] (\sh-\xa-0.15cm, \yy+\yg+1.95cm) -- (\sh-\xa-0.15cm, \yy+\yg+1.025cm);
\draw [decorate,decoration={snake, amplitude=0.05cm, segment length=0.2cm }] (\sh+\xa+0.15cm, \yy+\yg+1.05cm) -- (\sh+\xa+0.15cm, \yy+\yg+1.975cm);
\draw (\sh+\xa+0.20cm, \yy+\yg+1.0cm) -- (\sh+\xa+0.15cm, \yy+\yg+1.06cm);
\draw (\sh+\xa+0.15cm, \yy+\yg+1.96cm) -- (\sh+\xa+0.20cm, \yy+\yg+2.0cm);

\foreach \aaa in {1.0cm, 1.2cm, 1.4cm, 1.6cm, 1.8cm} {
\draw [domain=0:1, smooth, variable=\t] plot ({\sh-cos(\t*180.)*1.8cm}, {\yy+\yg+\t*0.2cm+\aaa-sin(\t*180.)*0.45cm}); }

\draw (\sh, \yy-1.5cm) node {\bf\large FIG. 3D};


\def\sh{0.0cm}     
\def\yy{-7.0cm}

\draw (\sh+\xa, \yy+\yg+1.0cm) -- (\sh+\xa, \yy+\yg-0.1cm) arc (90:-90:.07cm and .275cm) -- (\sh+\xa, \yy+\yj+0.1cm) arc (180:360:2.0cm and 0.5cm) -- (\sh-\xa, \yy+\yh+0.1cm)
      arc (-90:90:.07cm and .275cm) -- (\sh-\xa, \yy+\yg+1.0cm);
\draw (\sh+\xa, \yy+\yg-0.1cm) -- (\sh+\xa-0.3cm, \yy+\yg-0.1cm);
\draw (\sh+\xa, \yy+\yg-0.65cm) -- (\sh+\xa-0.3cm, \yy+\yg-0.65cm);
\draw (\sh+\xa-0.3cm, \yy+\yg-0.375cm) ellipse (.07cm and .275cm);

\draw (\sh-\xa, \yy+\yh+0.1cm) -- (\sh-\xa+0.3cm, \yy+\yh+0.1cm) arc (-90:90:.07cm and .275cm) -- (\sh-\xa, \yy+\yg-0.1cm);

\draw (\sh, \yy+\yg+1.0cm) ellipse (2.0 and 0.5);

\draw [fill=white, black!0] (\sh+\xa+0.2cm, \yy+\yg+2.0cm) arc (180:360:1.8 and 0.45) -- (\sh-\xa-0.2cm, \yy+\yg+1.0cm) arc (360:180:1.8 and 0.45) -- cycle;
\draw (\sh+\xa+0.2cm, \yy+\yg+2.0cm) arc (180:360:1.8 and 0.45);
\draw (\sh-\xa-0.2cm, \yy+\yg+1.0cm) arc (360:180:1.8 and 0.45);
\draw (\sh, \yy+\yg+2.0cm) ellipse (1.8 and 0.45);
\draw (\sh-1.0cm, \yy+\yg+1.8cm) -- (\sh-0.9cm, \yy+\yg+2.2cm) -- (\sh+0.9cm, \yy+\yg+2.2cm) -- (\sh+1.0cm, \yy+\yg+1.8cm) -- cycle;

\draw (\sh-\xa-0.20cm, \yy+\yg+1.0cm) -- (\sh-\xa-0.15cm, \yy+\yg+1.04cm);
\draw (\sh-\xa-0.20cm, \yy+\yg+2.0cm) -- (\sh-\xa-0.15cm, \yy+\yg+1.95cm);
\draw [decorate,decoration={snake, amplitude=0.05cm, segment length=0.2cm }] (\sh-\xa-0.15cm, \yy+\yg+1.95cm) -- (\sh-\xa-0.15cm, \yy+\yg+1.025cm);
\draw [decorate,decoration={snake, amplitude=0.05cm, segment length=0.2cm }] (\sh+\xa+0.15cm, \yy+\yg+1.05cm) -- (\sh+\xa+0.15cm, \yy+\yg+1.975cm);
\draw (\sh+\xa+0.20cm, \yy+\yg+1.0cm) -- (\sh+\xa+0.15cm, \yy+\yg+1.06cm);
\draw (\sh+\xa+0.15cm, \yy+\yg+1.96cm) -- (\sh+\xa+0.20cm, \yy+\yg+2.0cm);

\foreach \aaa in {1.0cm, 1.2cm, 1.4cm, 1.6cm, 1.8cm} {
\draw [domain=0:1, smooth, variable=\t] plot ({\sh-cos(\t*180.)*1.8cm}, {\yy+\yg+\t*0.2cm+\aaa-sin(\t*180.)*0.45cm}); }

\draw (\sh, \yy-1.5cm) node {\bf\large FIG. 3C};


\def\sh{12.5}     
\def\yy{5.0}

\draw (\sh, \yy) circle (1.0 and 0.25);
\draw [pattern=\paplug] (\sh, \yy) circle (0.5 and 0.125);
\draw (\sh+1.0, \yy) -- (\sh+1.0, \yy-1.0) -- (\sh+1.2, \yy-1.0) -- (\sh+1.2, \yy-1.1) -- (\sh+1.0, \yy-1.1) -- (\sh+1.0, \yy-2.0) -- (\sh+0.7, \yy-3.0) -- (\sh+0.7, \yy-8.0)
  arc (360:180:0.7 and 0.175) -- (\sh-0.7, \yy-3.0) -- (\sh-1.0, \yy-2.0) -- (\sh-1.0, \yy-1.1) -- (\sh-1.2, \yy-1.1) -- (\sh-1.2, \yy-1.0) -- (\sh-1.0, \yy-1.0) -- (\sh-1.0, \yy);

\draw (\sh-0.35, \yy-6.5) circle (0.2);
\draw (\sh+0.35, \yy-6.5) circle (0.2);

\draw (\sh+0.7, \yy-3.0) arc (360:180:0.7 and 0.175);

\draw (\sh+1.0, \yy-2.0) arc (360:180:1.0 and 0.25);

\draw (\sh, \yy-9.5) node {\bf\large FIG. 3E};

\def\locoffx{159}
\def\locoffy{451}


\refpointLR{\figA1   }{172}{759}{\sshh}{740}{342}{762} 
\refpointLR{\figA2   }{214}{700}{\sshh}{715}{302}{701} 
\refpointR {\figA3   }{182}{690}{\sshh}{690}           
\refpointR {\figA4   }{204}{665}{\sshh}{665}           
\refpointLR{\figA5   }{214}{640}{\sshh}{640}{363}{640} 
\refpointR {\figA6   }{180}{596}{\sshh}{615}           
\refpointR {\figA7   }{212}{579}{\sshh}{590}           
\refpointR {\figA8   }{213}{561}{\sshh}{565}           
\refpointR {\figA9   }{191}{558}{\sshh}{540}           
\refpointR {\figA{10}}{211}{548}{\sshh}{515}           
\refpointR {\figA{11}}{169}{550}{\sshh}{490}           
\refpointR {\figA{12}}{168}{517}{\sshh}{465}           
\refpointR {\figA{10}}{214}{470}{\sshh}{440}           
\refpointR {\figA9   }{166}{479}{\sshh}{415}           
\refpointLR{\figA6   }{158}{390}{\sshh}{390}{358}{390} 
\refpointLR{\figA7   }{212}{365}{\sshh}{365}{305}{365} 
\refpointLR{\figA{10}}{207}{340}{\sshh}{340}{312}{340} 
\refpointR {\figA{13}}{226}{320}{\sshh}{315}           

\refpointL {\figA{14}}{500}{515}{450}{480}             

\end{tikzpicture}

\bigskip
\begin{center}

\underline{\bf\large FIG. 3.}

\end{center}

\newpage

\noindent
\begin{tikzpicture}
  \def\plx { 4.0} 
  \def\plz { 0.2} 
  \def\plxx{ 4.0} 
  \def\plzz{ 3.0} 

  \def\yy{ 0.0} 

 \draw [line width=0.04cm, white, fill=white] (-\plx, \yy) -- (\plx, \yy) -- (\plx+\plxx, \yy+\plzz) -- (\plx+\plxx, \yy+\plzz+\plz) -- (-\plx+\plxx, \yy+\plzz+\plz) -- (-\plx, \yy+\plz) -- cycle;
 \draw [line width=0.04cm] (-\plx, \yy) -- (\plx, \yy) -- (\plx+\plxx, \yy+\plzz) -- (\plx+\plxx, \yy+\plzz+\plz) -- (-\plx+\plxx, \yy+\plzz+\plz) -- (-\plx, \yy+\plz) -- cycle;
 \draw [line width=0.04cm] (-\plx, \yy+\plz) -- (\plx, \yy+\plz) -- (\plx+\plxx, \yy+\plzz+\plz);
 \draw [line width=0.04cm] (\plx, \yy) -- (\plx, \yy+\plz);

  \def\yy{ 2.0} 
 
  \draw [line width = 0.04cm] (2.7+\plxx*0.5, \yy+\plzz*0.5+\plz) -- (2.7+\plxx*0.5, \plzz*0.5) arc (360:180:2.55 and 0.65) -- (-2.4+\plxx*0.5, \yy+\plzz*0.5+\plz);

 \def\mul{0.08}

 \draw [line width=0.2cm] (-\plx+\mul*\plxx*2.5, \plzz* \mul+\plz) -- (-\plx+\mul*\plxx*2.5, \yy+\plzz* \mul);
 \draw [line width=0.2cm] ( \plx-\mul*\plxx*0.5, \plzz* \mul+\plz) -- ( \plx-\mul*\plxx*0.5, \yy+\plzz* \mul);
 \draw [line width=0.2cm] (\plx-\mul*\plxx*2.5+\plxx, \plzz-\mul* \plzz+\plz) -- (\plx-\mul*\plxx*2.5+\plxx, \plzz-\mul* \plzz+\yy);

 \draw [line width=0.04cm, white, fill=white] (-\plx, \yy) -- (\plx, \yy) -- (\plx+\plxx, \yy+\plzz) -- (\plx+\plxx, \yy+\plzz+\plz) -- (-\plx+\plxx, \yy+\plzz+\plz) -- (-\plx, \yy+\plz) -- cycle;
 \draw [line width=0.04cm] (-\plx, \yy) -- (\plx, \yy) -- (\plx+\plxx, \yy+\plzz) -- (\plx+\plxx, \yy+\plzz+\plz) -- (-\plx+\plxx, \yy+\plzz+\plz) -- (-\plx, \yy+\plz) -- cycle;
 \draw [line width=0.04cm] (-\plx, \yy+\plz) -- (\plx, \yy+\plz) -- (\plx+\plxx, \yy+\plzz+\plz);
 \draw [line width=0.04cm] (\plx, \yy) -- (\plx, \yy+\plz);

 \draw [line width=0.2cm] (-\plx+\mul*\plxx*2.5, \yy+\plzz*\mul+\plz)            circle (0.2 and 0.05);
 \draw [line width=0.2cm] ( \plx-\mul*\plxx*0.5, \yy+\plzz*\mul+\plz)            circle (0.2 and 0.05);
 \draw [line width=0.2cm] (\plx-\mul*\plxx*2.5+\plxx, \yy+\plzz-\mul*\plzz+\plz) circle (0.2 and 0.05);
 \draw [line width=0.2cm] (-\plx+\mul*\plxx*0.5+\plxx, \yy+\plzz-\mul*\plzz+\plz) circle (0.2 and 0.05);

 \def\shsh{0.15}

\foreach \sh in {2.7, 2.4, 2.1, 1.8, 1.5, 1.2, 0.9}
{ \draw [line width = 0.03cm, dashdotted] (\sh+\plxx*0.5, \yy+\plzz*0.5+\plz) arc (360:180:{\sh-\shsh} and {\sh/4-\shsh/4}) arc (180:0:{\sh-\shsh*2} and {\sh/4-\shsh/2});
}

 \draw [line width = 0.04cm, dashdotted] (2.7+\plxx*0.5, \yy+\plzz*0.5+\plz) -- (2.7+\plxx*0.5, \plzz*0.5) arc (360:180:2.55 and 0.65) -- (-2.4+\plxx*0.5, \yy+\plzz*0.5+\plz);

 \draw [line width=0.07cm] (\plxx*0.5, \yy+\plzz*0.5+\plz) circle (0.56 and 0.14);
 
 \def\plk{0.3}

 \draw [line width=0.07cm, dashdotted] (\plxx*0.5, \plzz*0.5+\plz+\plk) circle (0.56 and 0.14);

 \draw [line width=0.04cm, dashdotted] (\plxx*0.5-0.56, \yy+\plzz*0.5+\plz) -- (\plxx*0.5-0.56, \plzz*0.5+\plz+\plk);
 \draw [line width=0.04cm, dashdotted] (\plxx*0.5+0.56, \yy+\plzz*0.5+\plz) -- (\plxx*0.5+0.56, \plzz*0.5+\plz+\plk);

 \draw [line width = 0.04cm, dashdotted] (\plxx*0.5+0.56, \plzz*0.5+\plz+\plk) -- (\plxx*0.5+0.56, \plzz*0.5+\plz) arc (360:180:0.56 and 0.14) -- (\plxx*0.5-0.56, \plzz*0.5+\plz+\plk);

 \draw [->, line width = 0.07cm] (\plxx*0.5, \plzz*0.5+5.0) -- (\plxx*0.5, \plzz*0.5+3.1);

\def\locoffx{188}
\def\locoffy{191}


\refpointR{\figM1}{245}{345}{275}{355} 
\refpointR{\figM2}{247}{297}{325}{355} 
\refpointR{\figM3}{371}{303}{460}{315} 
\refpointR{\figM4}{322}{255}{460}{285} 
\refpointR{\figM5}{260}{243}{460}{255} 
\refpointR{\figM6}{298}{215}{460}{225} 
\refpointR{\figM7}{302}{195}{460}{195} 

\end{tikzpicture}

\renewcommand{\figA}[1]{{\bf A.{#1}}\xspace}
\renewcommand{\figS}[1]{{\bf S.{#1}}\xspace}
\renewcommand{\figM}[1]{{\bf M.{#1}}\xspace}

\myskip
\underline{\bf FIG. 4}:
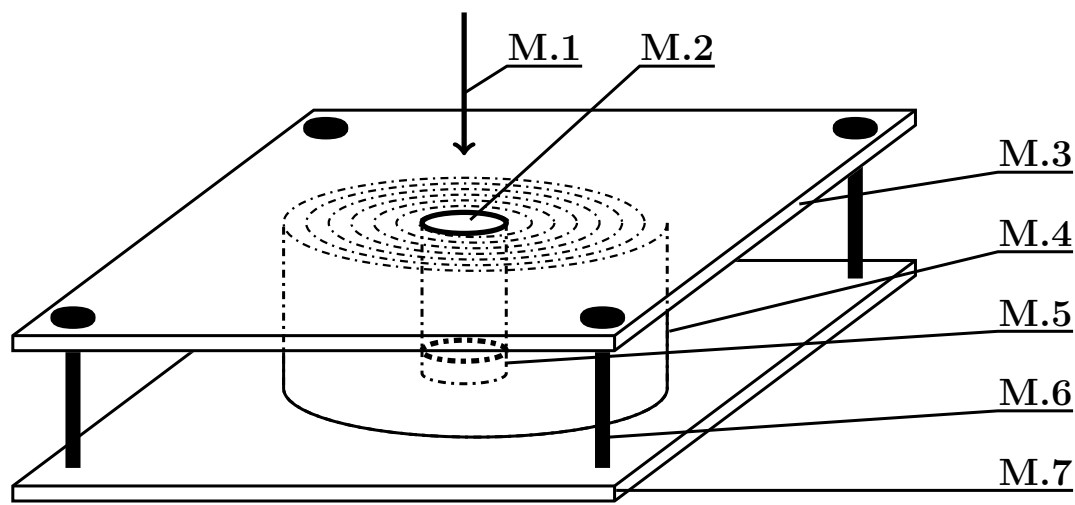
An apparatus for magnet recovery in in-situ NMR spectrometers.

\myskip
\figM1 Direction to place in-situ NMR spectrometer.

\myskip
\figM2 A hole into which the in-situ NMR spectrometer should be installed. 

\myskip
\figM3 Upper side of a device that prevents the coil \figM4 from moving up. 

\myskip
\figM4 The coil for generation of a permanent magnetic field. It consists of a foil of good conducted metal, preferably copper, with thin conductive resistant coverage.
The total amount of winding in this coil should be sufficient to generate a permanent magnetic field at least the same as the magnetic field that delivers magnets to
in-situ NMR spectrometers.

\myskip
Construction of the coil may be accomplished with one coil or several sections of coils so that the next section is connected to the previous section, all coils should have
winding in the same direction. If two sections are used, the end of the bottom section is situated inside the coil and is connected to the beginning of the top section,
which is situated inside the top section. Hence, two sections of coil assembly have both connections situated on outside components.
This coil should be connected over an electronic or mechanical switch to one or several capacitors, and/or super-capacitors, and/or batteries, and/or power supply units
connected in parallel, which should be capable of delivering enough current so that the coil is able to reach the permanent magnetic field to recover depleted magnets in
the in-situ NMR spectrometer. 

\myskip
\figM5 A supporting block at the bottom of the hole \figM2, on which the in-situ NMR spectrometer is situated during the recovery procedure. The height of this block should
be chosen to ensure that the maximal magnetic field area from the coil \figM4 is equal to the placement of permanent magnets in the in-situ NMR spectrometer. 

\myskip
\figM6 Mechanical joints between \figM3 and \figM7 that are strong enough to sustain the force between permanent magnets in the in-situ NMR spectrometer and the coil \figM4.

\myskip
\figM7 Bottom side of the assembly to prevent the in-situ NMR spectrometer from descending during magnetization.

\bigskip

\noindent
\begin{tikzpicture}

\newcommand{\mycoil}[8]{%
       \def\mcxb{#1}
       \def\mcyb{#2}
       \def\mcxe{#3}
       \def\mcye{#4}
       \def\mctol{#5}
       \def\mclen{7.0}
       \def\mcdow{2.3}
       \def\mcdx{#6}
       \def\mcdy{#7}
       \def\mcsh{0.4}

  \draw (\mcxb-\mctol, \mcyb-\mcdow-\mcsh) node {\bf\large #8};
  \draw (\mcxe+\mctol+2*\mcdx, \mcye-\mcdow+1*\mcdy-\mcsh) node {\bf\large #8};

  \draw[->, line width = 2.0, rounded corners=10pt  ] (\mcxb-\mctol, \mcyb-\mcdow) -- (\mcxb-0.5*\mctol, \mcyb-0.5*\mcdow);

  \draw[->, line width = 2.0, rounded corners=10pt ] (\mcxb-0.5*\mctol, \mcyb-0.5*\mcdow) --
     (\mcxb+0*\mcdx, \mcyb+0*\mcdy) -- (\mcxb+0*\mcdx, \mcyb+\mclen+0*\mcdy) -- (\mcxe+0*\mcdx, \mcye+\mclen+0*\mcdy) -- (\mcxe+0*\mcdx, \mcye+0*\mcdy) --
     (\mcxb+1*\mcdx, \mcyb+0*\mcdy) -- (\mcxb+1*\mcdx, \mcyb+\mclen+1*\mcdy) -- (\mcxe+1*\mcdx, \mcye+\mclen+1*\mcdy) -- (\mcxe+1*\mcdx, \mcye+1*\mcdy) --
     (\mcxb+2*\mcdx, \mcyb+1*\mcdy) -- (\mcxb+2*\mcdx, \mcyb+\mclen+2*\mcdy) -- (\mcxe+2*\mcdx, \mcye+\mclen+2*\mcdy) -- (\mcxe+2*\mcdx, \mcye+2*\mcdy) --
     (\mcxe+\mctol+2*\mcdx, \mcye-\mcdow+1*\mcdy);
}

\newcommand{\mcprint}[3]{%
\pgfmathsetmacro\mmcx{#1}
\pgfmathsetmacro\mmcy{#2}
  \draw (\mmcx, \mmcy) node {\bf\large #3};
}

\def\xa{0}
\def\ya{0}

\def\xb{7}
\def\yb{-2.0}

\def\xc{10}
\def\yc{2.0}

\pgfmathsetmacro\xd{\xa+\xc-\xb}
\pgfmathsetmacro\yd{\ya+\yc-\yb}

\def\tol{0.7}
\def\sh{0.3}
\def\sq{0.2}
\def\ss{0.1}
\def\len{7.0*1.15}

\mycoil{\xa    }{\ya    }{\xb    }{\yb    }{\tol}{\sh}{ \sh}{1}
\mycoil{\xb+\ss}{\yb+\ss}{\xc+\ss}{\yc+\ss}{\tol}{\sh}{-\sh}{2}
\mycoil{\xc    }{\yc    }{\xd    }{\yd    }{\tol}{\sh}{ \sh}{3}
\mycoil{\xd+\ss}{\yd+\ss}{\xa+\ss}{\ya+\ss}{\tol}{\sh}{-\sh}{4}

\mycoil{\xa+\sq}{\ya+\sq}{\xc+\sq}{\yc+\sq}{0}{\sh}{\sh}{5}
\mycoil{\xb+\sq}{\yb+\sq}{\xd+\sq}{\yd+\sq}{0}{\sh}{\sh}{6}

\mcprint{(\xa+\xb)*0.5+1.0}{(\ya+\yb)*0.5+\len-0.2}{1}
\mcprint{(\xb+\xc)*0.5    }{(\yb+\yc)*0.5+\len-0.5}{2}
\mcprint{(\xc+\xd)*0.5+1.0}{(\yc+\yd)*0.5+\len-0.2}{3}
\mcprint{(\xd+\xa)*0.5    }{(\yd+\ya)*0.5+\len-0.5}{4}
\mcprint{(\xa+\xc)*0.5-0.8}{(\ya+\yc)*0.5+\len-0.2}{5}
\mcprint{(\xb+\xd)*0.5+0.5}{(\yb+\yd)*0.5+\len+0.7}{6}

\end{tikzpicture}

\bigskip

\noindent
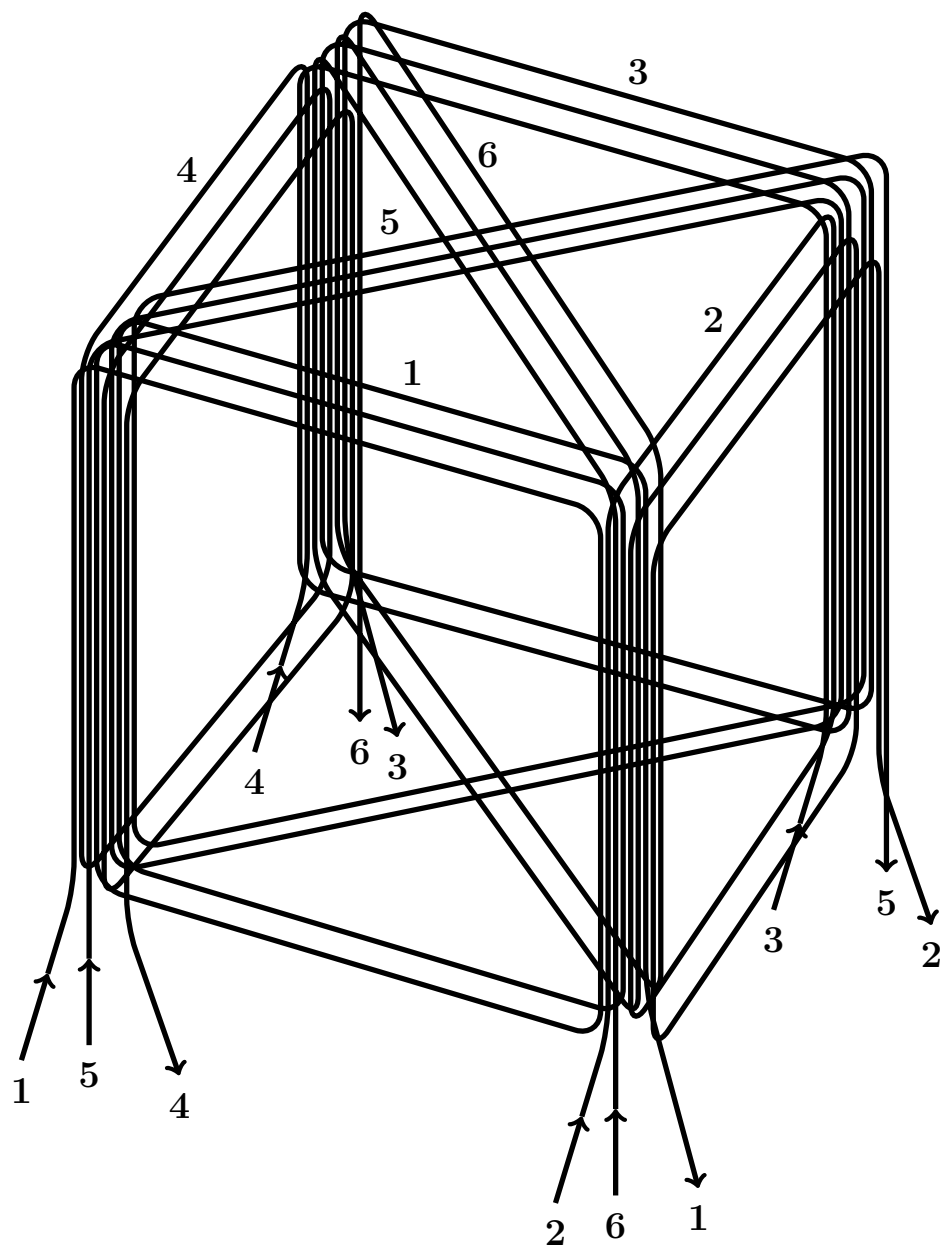
\underline{\bf FIG. 5.} Six coils situated on the edges of parallelepiped (three-dimensional figure formed by six parallelograms). The total number of turns for
each coil may be two or more. The numbers of turns on coils within each subset \{1, 2, 3, 4\} and \{5, 6\} are equal, i.e. coils  \{1, 2, 3, 4\} must be the same and
coils \{5, 6\} must be the sabe also, but coils  \{1, 2, 3, 4\} can differ from coils \{5, 6\}. The optimal number of turns in each subset depends on the dimensions of
the device, the magnetic field's strength, and the electronics used. Depending on the embodiment, each coil subset may be comprised of transmitting and/or receiving coils.

\newpage

{\bf\large

\noindent
APPENDIX: MINIMIZATION ALGORITHM

}

\bigskip

\myskip
The minimization function in the problem (1) is independent on the $t$ variable, hence it is sufficient to solve:
$$
\min_{x, y} \sum_{j=1}^J ||A_j - y (q_j + x)^*||_2^2,
$$
where
\myskip
\begin{itemize}
\item $x \in \C^N$ and $y \in \C^I$ are unknowns that refer to $p_n(t)$ and $\epsilon_j(t)$,
\item $A_j \in \C^{I \times N} = \{ a_{inj} \}$ refers to to $u_{lnj}$,
\item $q_j \in \C^N$ refers to $q_{ln}$,
\end{itemize}
for each given $t$.
\myskip
The following objects should be computed step-by-step:
$$
P = \sum_{j=1}^J A_j,
$$
then compute the SVD \cite{SVD} of $P = U D V^*, ~~ U \in \C^{I \times R}, ~~ U^* U = I, ~~ V \in \C^{N \times R}, V^* V = I, ~~
D = {\rm diag}(d), \in \R^{R \times R} ~~ d = (d_1, \dots, d_R)^T, ~~ d_1 \ge \dots \ge d_R, ~~ R = \min(I, N)$
and afterwards compute
$$
a = \sum_{j=1}^J ||q_j||_2^2, ~~
b = V^* \sum_{j=1}^J q_j = (b_1,\dots,b_R)^T, ~~
c = U^* \sum_{j=1}^J A_j q_j = (c_1,\dots,c_R)^T.
$$
\myskip
Find by the bisection method a root of
$$
J a - ||b||_2^2 = \sum_r \frac{|J c_r - b_r d_r|^2} {\gamma J - d_r^2}
$$
by the $\gamma$ variable in an open interval $\gamma > d_1^2/J$ with upper bound taken as the computer precision and compute
$$
\forall r=1,\dots, R:
z_r = \frac{ \gamma b_r - d_r c_r } { d_r^2 - J \gamma }, ~~
x = Vz, ~~ y = \gamma ||Dz+c||_2^2 (P x + U c).
$$
\myskip
There are many special cases, for example if one or several isotopes are missing in the investigated substances. In the case where only 2 input coils are considered
($J=2$) and the first coil reads reference spectra only, the solution reads as: 
$$y = A_1 q_1/||q_1||_2^2, ~~ x = \frac{||q_1||_2^2}{||A_1 q_1||_2^2} A_2^* A_1 q_1 - q_2.$$
Additionally, in the case of one or several entries in $A_j$ arrays being missing, the sparse multi-dimensional decomposition algorithm described in the chapter ``sparse three-way decomposition'' of \cite{IIV2002} should be used. 

\end{document}